\documentclass[conference]{IEEEtran}
\usepackage[tbtags]{amsmath}
\usepackage{amssymb}
\usepackage{verbatim}
\usepackage{amsxtra}
\usepackage{enumerate}
\usepackage{amsfonts}
\usepackage{color}
\usepackage{cite}
\usepackage{epsfig, graphicx, psfrag}
\usepackage{units}
\usepackage[table]{xcolor}   
\usepackage{calc}
\usepackage{amsfonts,amssymb,graphics,psfrag,subfigure,epsfig,graphics}
\usepackage{epstopdf}
\usepackage[mathcal]{euscript}
\usepackage{multirow}
\usepackage{arydshln}
\usepackage{mathtools}
\mathtoolsset{showonlyrefs=true}
\usepackage{url}
\usepackage{caption}
\newtheorem{thm}{Theorem}

\newtheorem{remark}{Remark}

\newtheorem{defn}{Definition}

\providecommand{\thmref}[1]{Theorem~\ref{#1}}

\providecommand{\secref}[1]{Section~\ref{#1}}

\providecommand{\tableref}[1]{Table~\ref{#1}}

\newcommand{\bm}[1]{\mbox{\boldmath{$#1$}}}

\newcommand{\mA}{\mathcal{A}}

\newcommand{\mV}{\mathcal{V}}

\newcommand{\mU}{\mathcal{U}}

\newcommand{\Comment}[1]{}
\newcommand{\old}[1]{}
\newcommand{\rem}[1]{}

\newcommand{\tG}{\tilde G}
\newcommand{\tQ}{\tilde Q}

\newcommand{\by}{\bm y}

\newcommand{\bx}{{\bm x}}

\newcommand{\bz}{{\bm z}}

\providecommand{\bx}{{\bf x}}
\providecommand{\by}{{\bf y}}

\providecommand{\bz}{{\bf z}}

\providecommand{\comment}[1]{}

\newcommand{\beqn}[1]{\begin{eqnarray}\label{#1}}
\newcommand{\eeqn}{\end{eqnarray}}
\newcommand{\beq}[1]{\begin{equation}\label{#1}}
\newcommand{\eeq}{\end{equation}}

\newcommand{\tby}{\tilde {\bm y}}
\newcommand{\tbx}{\tilde {\bm x}}

\providecommand{\submat}[3]{#1 \left\lfloor #2, \, #3 \right\rceil}

\newcommand{\blath}[1]{\mbox{#1-th}}

\newcommand{\CC}{C_\bx}

\newcommand{\UUU}{\mathcal{U}}
\newcommand{\VVV}{\mathcal{V}}

\newcommand{\OOO}{\mathcal{O}}
\newcommand{\AAA}{\mathcal{A}}

\newcommand{\smallr}{r}
\newcommand{\bigT}{T}
\newcommand{\bigR}{R}

\DeclareMathAlphabet{\mathbsf}{OT1}{cmss}{bx}{n}
\DeclareMathAlphabet{\mathssf}{OT1}{cmss}{m}{sl}
\DeclareMathAlphabet{\mathcsf}{OT1}{cmss}{sbc}{n}

\DeclareSymbolFont{bsfletters}{OT1}{cmss}{bx}{n}
\DeclareSymbolFont{ssfletters}{OT1}{cmss}{m}{n}
\DeclareMathSymbol{\bsfGamma}{0}{bsfletters}{'000}
\DeclareMathSymbol{\ssfGamma}{0}{ssfletters}{'000}
\DeclareMathSymbol{\bsfDelta}{0}{bsfletters}{'001}
\DeclareMathSymbol{\ssfDelta}{0}{ssfletters}{'001}
\DeclareMathSymbol{\bsfTheta}{0}{bsfletters}{'002}
\DeclareMathSymbol{\ssfTheta}{0}{ssfletters}{'002}
\DeclareMathSymbol{\bsfLambda}{0}{bsfletters}{'003}
\DeclareMathSymbol{\ssfLambda}{0}{ssfletters}{'003}
\DeclareMathSymbol{\bsfXi}{0}{bsfletters}{'004}
\DeclareMathSymbol{\ssfXi}{0}{ssfletters}{'004}
\DeclareMathSymbol{\bsfPi}{0}{bsfletters}{'005}
\DeclareMathSymbol{\ssfPi}{0}{ssfletters}{'005}
\DeclareMathSymbol{\bsfSigma}{0}{bsfletters}{'006}
\DeclareMathSymbol{\ssfSigma}{0}{ssfletters}{'006}
\DeclareMathSymbol{\bsfUpsilon}{0}{bsfletters}{'007}
\DeclareMathSymbol{\ssfUpsilon}{0}{ssfletters}{'007}
\DeclareMathSymbol{\bsfPhi}{0}{bsfletters}{'010}
\DeclareMathSymbol{\ssfPhi}{0}{ssfletters}{'010}
\DeclareMathSymbol{\bsfPsi}{0}{bsfletters}{'011}
\DeclareMathSymbol{\ssfPsi}{0}{ssfletters}{'011}
\DeclareMathSymbol{\bsfOmega}{0}{bsfletters}{'012}
\DeclareMathSymbol{\ssfOmega}{0}{ssfletters}{'012}
\newcommand{\UUUU}[2]{{  \left(\UUU_{#1}^{\left({#2}\right)}\right)^\dagger    }}
\newcommand{\Ul}[2]{{  \left(U_{#1}^{\left({#2}\right)}\right)^\dagger    }}
\newcommand{\VVVV}[1]{{  \VVV^{\left({#1}\right)}    }}
\newcommand{\Vl}[1]{{  V^{\left({#1}\right)}    }}
\providecommand{\subm}[2]{\left\lfloor #1, \, #2 \right\rceil}

\providecommand{\emb}[2]{I_{#1}^{#2}}

\IEEEoverridecommandlockouts 

\begin{document}

\title{Space--Time MIMO Multicasting}

\author{
    Idan Livni, Anatoly Khina, Ayal Hitron, Uri Erez$^*$ \\
    {\em \{idanlivn,anatolyk,ayal,uri\}@eng.tau.ac.il} \\
    EE-Systems Dept., Tel Aviv University, Tel Aviv, Israel
    \thanks{$^*$ This work was supported in part by the ISF under Grant No.~1557/12.}
}
\maketitle


\begin{abstract}
    Multicasting is the general method of conveying
    the same information to multiple users over a broadcast channel.
    In this work, the Gaussian MIMO broadcast channel is considered, with multiple users and any number of antennas at each node.
    A ``closed loop'' scenario is assumed, for which a practical capacity-achieving multicast scheme is constructed.
    In the proposed scheme, linear modulation is carried over time and space together, which allows
    to transform the problem into that of transmission over parallel scalar sub-channels,
    the gains of which are equal, except for a fraction of sub-channels that vanishes with the number of time slots used.
    Over these sub-channels, off-the-shelf fixed-rate AWGN codes can be used to approach capacity.
\end{abstract}

\section{Introduction}
\label{sec:intro}

A recurring theme in digital communications is the use of a standard ``off-the-shelf'' coding module in
combination with appropriate linear pre/post processing which is tailored to the specific channel model. Such
methods are appealing due to their low complexity of implementation as well as conceptually, since the task
of coding and modulation are effectively decoupled.

Underlying such decoupling schemes is the existence of a diagonalization transformation. For
time-invariant
scalar systems, this is possible via the Fourier transform. The singular-value decomposition (SVD)
plays a similar role for Gaussian multiple-input multiple-output (MIMO) channels.
In recent years, as coding and decoding for single-user scalar channels has reached a mature stage, research
effort has shifted to tackling the more ambitious goal of efficient multi-user (and multi-antenna)
communication networks.

Extension of the decoupling approach which is at the heart of single-user scalar systems to a multiple-user
MIMO network requires, however, overcoming a major hurdle: (unitary) simultaneous diagonalization is in
general not possible.\footnote{Even if all matrices are diagonal, constructing a practical capacity-achieving scheme is hard,
since using scalar coding over the resulting parallel channels is limited to working w.r.t.\ the minimal gain over each sub-channel.}

Hence, different practical approaches were proposed over the years for the problem of multicasting over Gaussian MIMO broadcast channels.
However, none of these approaches is capacity achieving in general, even for simple cases. To illustrate this, consider the following simple three-user example:
\begin{align*}
    \by_i = H_i \bx + \bz_i \,, \qquad i=1,2,3 \,,
\end{align*}

\noindent where $\bz_i$ are white Gaussian noises with unit power (for each element), $\bx$ is the channel vector subject to an average power constraint $P$,
and $H_i$ are the complex-valued channel matrices
\begin{align}
\label{eq:example_ChannelMatrices}
    H_1 = \left(
            \begin{array}{cc}
              \alpha & 0 \\
              0 & \alpha \\
            \end{array}
          \right)
    , \:\:
    H_2 = \left(
            \begin{array}{cc}
              \beta & 0 \\
            \end{array}
          \right)
    , \:\:
    H_3 = \left(
            \begin{array}{cc}
              0 & \beta \\
            \end{array}
          \right)
\end{align}

\noindent where $\alpha$ and $\beta$ are chosen such that the (individual) capacities of all channels are equal, viz.\
\begin{align}
\label{eq:example_IndividualCapacities}
  C^\text{p2p} \triangleq
  2 \log(1 + |\alpha|^2 P/2) = \log(1 + |\beta|^2 P) \,.
\end{align}

This example models a ``near--far'' scenario in which the two ``near'' users have one antenna each,
where each receives a different transmit antenna stream,
whereas the ``far'' user is equipped with two antennas to compensate for the distance attenuation,
such that the capacities of all channels are equal.

A na\"ive approach would be to use Vertical Bell-Laboratories Space--Time coding (V-BLAST) \cite{Wolniansky_V-BLAST} or generalized decision feedback equalization (GDFE) \cite{CioffiForneyGDFE}.
This approach is based upon the QR decomposition, in which the output of the channel is multiplied by a unitary matrix, resulting in an effective triangular channel matrix.
In the case of the example, however, repetition coding across the antennas must be used, to convey the (common) message to users 2 and 3.
This, in turn, implies that the far user (user 1) cannot enjoy any of its multiplexing gain (which is equal to two in this case),
due to the repetition that is carried across the two transmit antennas.
Again, in the high SNR regime, this suggests a loss of approximately half of the optimal achievable rate.
Moreover, for this example the ``max--min beamforming'' technique (see, e.g., \cite{Zoltowski_MaxMinBearmforming} and references therein)
reduces to this scheme as well, meaning that it suffers the same losses in performance.

Another approach considered in the literature for this problem is that of using a ``pure open-loop'' approach,
namely Alamouti \cite{Alamouti}~--- for the two-transmit antenna case, and space--time coding \cite{TarokhSheshadriCalderbank_STC}~--- for more.
The performance of these schemes does not depend on the number of receivers.
However, this universality comes at the price of a substantial rate loss for MIMO channels having several receive antennas, as these schemes use only a single stream, thus failing to achieve the multiplexing gain offered by the MIMO channel.\footnote{Moreover, for more than two transmit antennas, the space--time codes of \cite{TarokhSheshadriCalderbank_STC} attain strictly less than one degree of freedom.}

Finally note that time/frequency sharing suggest a great loss in performance (up to two thirds of the capacity in this case).

In general, to the best of our knowledge known practical schemes are limited to the smallest number of degrees of freedom (``multiplexing gain``) of the different users,
or alternatively incorporate time- or frequency-sharing, which again lose degrees of freedom.
Thus, these schemes achieve only a fraction of the available degrees of freedom.

In this work we develop a scheme that achieves the degrees of freedom of each individual channel,
by enabling the transmission of several streams as is done in \mbox{V-BLAST}/GDFE in the point-to-point case.
This is done by designing a special space--time coding structure that is tailored for the specific channel matrices,
where the number of channel uses that is needed to be jointly processed depends on the number of users in the system.
However, in contrast to the open-loop space--time coding structures, which strive for an ``orthogonal design'' structure (see, e.g., \cite{TarokhSheshadriCalderbank_STC}),
the space--time structure presented in this work results in triangular forms, similar to V-BLAST/GDFE, but having \emph{equal diagonals},
which suggests in turn, the optimality of the scheme.
This gives rise to effective parallel scalar additive white Gaussian noise (AWGN) channels,
over which standard codes can be used to approach capacity.
Thus, the proposed scheme could be thought of as an interpolation between the open-loop space--time coding technique and the point-to-point V-BLAST one.

\section{Channel Model}

The $K$-user Gaussian MIMO broadcast channel consists of one transmit and $K$ receive nodes, where each received signal is related to the transmitted signal through a MIMO link:
\beq{MIMO-BC}
  \by_i = H_i \bx + \bz_i \,, \qquad i = 1,\dots,K \,,
\eeq

\noindent where $\bx$ is the channel input of dimensions $n_t \times 1$ subject to an average power constraint $P$;\footnote{Alternatively, one can consider an input covariance constraint
$E\left[ \bx \bx^\dagger \right] \preceq C$,
where by $C_1 \preceq C_2$ we mean that
$\left(C_2 - C_1\right)$ is positive semi-definite.}
$\by_i$ is the channel output vector of receiver $i$ ($i=1,\dots,K$) of dimensions
$n_r^{(i)} \times 1$; $H_i$ is the channel matrix to user $i$ of dimensions $n_r^{(i)} \times n_t$ and $\bz_i$ is an additive circularly-symmetric Gaussian noise vector of dimensions $n_r^{(i)} \times 1$, where, without loss of generality, we assume that the noise elements are mutually independent and identically distributed with unit power.

The aim of the transmitter is to multicast the same (common) message to all the receivers.
The capacity of this scenario is long known to equal the (worst-case) capacity of the
compound channel (see, e.g., \cite{BlackwellBreimanThomasian59}), with the compound parameter being the channel matrix index:
\begin{align}
\label{eq:MIMO_BC_capacity}
    C \left( \left\{ H_i \right\}_{i=1}^K, P \right) = \max_{C_\bx} \min_{i=1,\dots,K} I(H_i,C_\bx) \,,
\end{align}

\noindent where $I(H_i,C_\bx)$ is the mutual information between the channel input $\bx$ and the channel output
$\by_i$, obtained by taking $\bx$ to be Gaussian with covariance matrix $C_\bx$:
\begin{align} \label{MIMO_MI}
  I(H,C_\bx) \triangleq \log \det \left( I + H_i \CC H_i^\dagger \right) \,,
\end{align}

\noindent and the maximization is carried over all admissible input covariance matrices $C_\bx$, satisfying the power constraint.

\section{Background}
\label{sec:Background_Network_Modulation}

In this section we recall the transmission and receiving scheme for the single- and two-user cases, and explain how
this scheme can be generalized to the multi-user case.

\subsection{Unitary Matrix Triangularization}

The proposed scheme in this section is based on several forms of matrix decompositions, one of which is the \emph{geometric mean decomposition} (GMD) \cite{GMD}. For simplicity, we will only consider the decomposition of \emph{square} matrices throughout this work. As we show in the sequel, this does not pose any restriction on the communication problem addressed. The GMD  \cite{GMD} of a square complex invertible matrix $A$ is given by:

    \begin{align} \label{eq:gmd}
        A &= U \bigT V^\dagger \,,
    \end{align}

\noindent where $U$, $V$ are unitary matrices, and $\bigT$ is an upper-triangular matrix
such that all its diagonal values equal
to the geometric mean of the singular values of $A$, which is real and positive.

Building on the GMD, the following decomposition, which will be referred to as Joint Equi-diagonal Triangularization (JET),  was introduced in \cite{STUD:SP}. Let  $A_1$ and $A_2$ be two invertible complex matrices of dimensions $n \times n$
such that \mbox{$|\det (A_1)| = |\det (A_2)|$}.
Then, the joint triangularization of $A_1$ and $A_2$ is given by:

\begin{align}
\begin{aligned}
  A_1 &= U_1 \bigR_1 V^\dagger \\
  A_2 &= U_2 \bigR_2 V^\dagger \,,
\end{aligned}
\label{eq:jet2}
\end{align}

\noindent where $U_1,U_2,V$ are $n \times n$ unitary matrices,     and $\bigR_1,\bigR_2$ are upper-triangular
 $n \times n$ matrices with \emph{the same} real-valued, non-negative diagonal values, namely,
$$
    \left[ \bigR_1 \right]_{ii} = \left[ \bigR_2 \right]_{ii} \,\quad \forall i=1,\ldots,n \,.
$$

\subsection{Point-to-Point MIMO Scheme via Matrix Triangularization}
\label{ss:p2p_scheme}
We now review the transmission scheme known as the uniform channel decomposition (UCD) \cite{UCD}, which is in turn based upon the derivation
of the MMSE version of Vertical Bell-Laboratories Space--Time coding (V-BLAST), see, e.g., \cite{HassibiVBLAST}.
Later in the paper we take the triangularization to be one which is simultaneously good for several users.

Define the following \emph{augmented matrix}:\footnote{$\CC^{1/2}$ is any matrix $B$ satisfying: $B B^\dagger = \CC$, and can be found, e.g., via the Cholesky decomposition.}

\begin{align*}
	\tG \triangleq
	\left(\begin{array}{c}
		  H \CC^{1/2} \\ I_{n_t}
	      \end{array}
	\right)\,,
\end{align*}
where $I_{n_t}$ is the $n_t \times n_t$ identity matrix.

Next, the matrix $\tG$ is transformed into a square matrix, by means of the QR decomposition:

\begin{align*}
 \tG = Q G \,,
\end{align*}
where $Q$ is an $(n_r+n_t) \times n_t$ matrix with orthonormal columns and $G$ is an $n_t \times n_t$ upper-triangular matrix with real-valued positive diagonal elements. Now the matrix G is decomposed according to the GMD:
\begin{align}
\label{eq:Ggmd}
    G &= U \bigT V^\dagger \,,
\end{align}
where $T$ is upper-triangular whose diagonal values are equal to $\sqrt[n_t]{\det{(G)}}$, 
and $\sqrt[n_t]{\det{(G)}} - 1$ is the effective signal-to-noise ratio of the scalar sub-channels..

The  transmission scheme is as follows:
\begin{enumerate}
\item
  Construct $n_t$ codewords of equal rates for a scalar AWGN channel of signal-to-noise ratio (SNR) $\sqrt[n_t]{\det{(G)}}-1$.

\item
  In each channel use, an $n_t$-length vector $\tbx$ is formed using one sample from each codebook. The transmitted vector $\bx$ is then obtained using the following precoder:
  \begin{align} \label{eq:bx_tilde_to_bx}
	  \bx = \CC^{1/2} V \tbx \,.
  \end{align}

\item
  The receiver calculates
  \begin{align} \label{eq:scheme_receiver}
	  \tby = U^\dagger \tQ^\dagger \by \,,
  \end{align}
  where $\tQ$ consists of the first $n_t$ rows of $Q$.

\item
  Finally, the codebooks are decoded using successive interference cancellation, starting from the \blath{$n_t$}  codeword and ending with the first one: The \blath{$n_t$} codeword is decoded first, using the \blath{$n_t$} element of $\tby$,
  treating the other codewords as AWGN. The  effect of the \blath{$n_t$} element of $\tbx$ is then subtracted out from the remaining elements of $\tby$. Next, the \blath{$(n_t-1)$} codeword is decoded, using the \blath{$(n_t-1)$} element of $\tby$ --- and so forth.
\end{enumerate}
The optimality of this scheme, i.e., that it is capacity achieving, was proved in~\cite[Sec.~IV]{STUD:SP}.

\section{MIMO Multicast Scheme}
\label{ss:mimo_multicasting_scheme}

The scheme of \secref{ss:p2p_scheme} can be generalized to the $K$-user case in a straightforward manner.\footnote{A similar scheme for the two-user MIMO multicast case was  proposed in \cite{STUD:SP}, where JET \eqref{eq:jet2} was used, implying the need of using scalar codebooks of different rates.}
However, in order to approach the capacity \eqref{eq:MIMO_BC_capacity}, using the same scalar codebook over all scalar sub-channel, as in \secref{ss:p2p_scheme}, the existence of a joint unitary matrix decomposition of the form
\begin{align}
\label{eq:PerfectK-GMD}
  A_i &= U_i T_i V^\dagger \,, & i=1,\dots,K \,,
\end{align}
is required, assuming $A_i$ are square invertible matrices with equal determinants (up to phase), of dimensions $n \times n$,
where $U_i$ are unitary matrices (corresponding to operations performed at the receivers),
$V$ is unitary as well (corresponding to an operation performed by the transmitter) and $T_i$
are \emph{upper-triangular} matrices with \emph{constant diagonals}.
Unfortunately, such a decomposition does not exist in general for more than one matrix since there are not enough degrees of freedom offered by the unitary matrices, as the unitary matrix on the right, $V$,
is the same for all decomposed matrices $\{A_i\}$ (corresponding to the common operation carried at the transmitter); for more details see \cite{JET:SeveralUsersISIT11}.
To overcome this problem, in order to gain more degrees of freedom,
we propose to utilize multiple channel uses of the same channel realization and process them together.
The idea of mixing the same symbols between multiple channel uses has much in common with space--time codes \cite{Alamouti,TarokhSheshadriCalderbank_STC}.
In the next section we show how, using this idea, nearly optimal joint triangularization may be obtained.

\section{Space--Time Triangularization}
\label{s:space_time_triangularization}

We now show how to utilize a space--time structure in order to obtain nearly-optimal joint triangularization of $K$ matrices,
such that the resulting triangular matrices have constant diagonals, up to a small portion of the diagonal extreme elements.
The resulting scheme becomes asymptotically optimal for large values of $N$, where $N$ is the number of channel uses grouped together for the purpose of joint decomposition. This result is stated in the following theorem.

\begin{thm}[Nearly-Optimal $K$-GMD]
\label{thm:n_n_asymptotical}
  Let $A_1,\dots,A_K$ be complex-valued $n \times n$ matrices satisfying $|\det(A_i)|=1$, and $N > n^{K-1}$.
  Define the following $nN \times nN$ extended matrices:
  \begin{align}
  \label{eq:extended_n}
    \AAA_i = \left(
    \begin{array}{ccccc}
      A_i & 0 & 0 & \cdots & 0 \\
      0 & A_i & 0 & \cdots & 0 \\
      \vdots & \cdots & \vdots & \ddots & \vdots \\
      0 & 0 & 0  & \cdots & A_i
    \end{array}
    \right)
  \,, \quad i=1,\dots,K \,.
  \end{align}
  Then there exist matrices $\UUU_1,\dots,\UUU_K,\VVV$, all of dimensions $nN \times n \big(N - ( n^{K-1} - 1 ) \big)$, with orthonormal columns, such that:
  \begin{align*}
    \UUU_i^\dagger \AAA_i \VVV = \left( \begin{array}{ccccc}
		    1 & * & \cdots & * & * \\
		    0 & 1 & \cdots & * & *\\
		    \vdots & \vdots & \ddots & \vdots & \vdots \\
		    0 & 0 & \cdots & 1 & * \\
		    0 & 0 & \cdots & 0 & 1
		\end{array} \right) \,,
    \qquad i=1, \dots, K \,,
  \end{align*}
  where $*$ represents some value (which may differ within each matrix as well as between different ones).
\end{thm}

By using this decomposition, the same scheme as in \secref{sec:Background_Network_Modulation} can be employed,
such that the $N$ channel uses are effectively transformed into \mbox{$n \big(N - ( n^{K-1} - 1 ) \big)$} equal-rate scalar AWGN channels.
The sum of the capacities of these channels tends to the capacity of the original channel for large values of $N$,
where the only loss~comes from edge effects (truncation of the extreme $\big( n^{K} - n \big)$ elements).

\begin{remark}
    For the case where the matrices have non-equal determinants, the $K$-GMD \thmref{thm:n_n_asymptotical} results in $K$ triangular matrices, each with a constant diagonal with entries that are equal to $\sqrt[n]{|\det{(A_i)}|}$.
\end{remark}

\begin{remark}
  It was shown in \cite[Lemma~1]{JET:SeveralUsersISIT11} that $K$-GMD is equivalent to $(K+1)$-JET.
  Hence, nearly-optimal $(K+1)$-JET can be obtained with the same parameters as in \thmref{thm:n_n_asymptotical}.
\end{remark}


We first present the tools used in the construction. We then demonstrate the construction for the special case of $2 \times 2$ matrices
$K=3$ users and $N=4$ augmentations.
The general case (utilizing the same tools) is given, as Matlab and Python codes in \cite{MATLAB,PYTHON}.

\begin{defn}
  Let $A$ and $B$ be matrices of dimensions $n \times m$ and $2 \times 2$, respectively.
  We define the operation of ``extraction'' of indices $i$ and $j$ from $A$ by:

  \begin{align*}
    \submat{A}{i}{j} \triangleq   \left(
				\begin{array}{cc}
				  A_{ii} & A_{ji} \\
				  A_{ji} & A_{jj}
				\end{array}
				\right)
    \,,
  \end{align*}

    \noindent where $\subm{i}{j} \triangleq \left\{ (i,j), (i,i), (j,i), (j,j) \right\}$.

    We further define the ``embedding'' operation $\emb{n}{B}\left( \bigcup_i \subm{m_i}{n_i} \right)$
    as the replacement of the elements in the identity matrix $I_n$
    in the index-pairs contained in
    $\subm{m_1}{n_1}\subm{m_2}{n_2}\subm{m_3}{n_3},\dots$,\footnote{By $\subm{i}{j}\subm{k}{l}$ we denote $\subm{i}{j} \cup \subm{k}{l}$.}
    with the elements contained in $B$,
    where index overlap is forbidden, i.e., all the indices $\left\{ m_i \right\} \cup \left\{ n_i \right\}$ are unique.
    For example, the embedding $\emb{4}{B}\left(\subm{1}{3}\subm{2}{4}\right)$ of
    \begin{align*}
      B = \left( \begin{array}{cc}
    	      11 & 2 \\
    	      3  & 4
    	     \end{array}
          \right)
    \end{align*}
    into the four-dimensional identity matrix $I_4$ is

      \begin{align*}
        \left(
        \begin{array}{cccc}
          \cellcolor[gray]{0.8} 11 & 0 & \cellcolor[gray]{0.8} 2 & 0 \\
          0  & \cellcolor[gray]{0.5} 11 & 0 & \cellcolor[gray]{0.5} 2 \\
          \cellcolor[gray]{0.8} 3  & 0 & \cellcolor[gray]{0.8} 4 & 0 \\
          0  & \cellcolor[gray]{0.5} 3 & 0 & \cellcolor[gray]{0.5} 4 \\
        \end{array}
        \right)
      \,.
      \end{align*}
\end{defn}

\noindent
\emph{ \underline{ \textbf{Algorithm for \boldmath{$n=2, K=3, N=4$}:} } }
  \noindent
  Denote by $\left\{ \mA_i \right\}$ the augmented matrices corresponding to $N = 4$ channel uses.

  \underline{Stage 1:}
  Start by applying a 1-GMD for each block (corresponding to a single channel use) of the first matrix $A_1$:

\begin{align}
\label{eq:stage_a}
	\Ul11 A_1 \Vl1 = \left( \begin{array}{cc}
	                     \small1 & * \\ 0 & \small1
	                 \end{array} \right)
	\,,
\end{align}

\noindent which corresponds, in turn, to applying the following extended unitary matrices

\begin{align*}
    \UUUU11 &\triangleq \emb{8}{\Ul11}\left(\subm{1}{2}\subm{3}{4}\subm{5}{6}\subm{7}{8}\right)\\
    \VVVV1 &\triangleq \emb{8}{\Vl1}\left(\subm{1}{2}\subm{3}{4}\subm{5}{6}\subm{7}{8}\right)
    \,.
\end{align*}

\noindent and results in the extended triangular matrix

  \begin{align*}
      &T_1^{(1)} = \UUUU11\AAA_1 \VVVV1\\
      &\quad = \left(
	\begin{array}{ccccccccc} \cline{1-2}
		\multicolumn{1}{|c}{\small1} & \multicolumn{1}{c|}{*} & 0 & 0 & 0 & 0 & 0 & 0 \\
		\multicolumn{1}{|c}{0}    &  \multicolumn{1}{c|}{\cellcolor[gray]{0.8} \small1} & 0 & 0 & \multicolumn{1}{c}{\cellcolor[gray]{0.8} \small0} & 0 & 0 & 0 \\ \cline{1-4}
		  0 & 0 & \multicolumn{1}{|c}{\small1} & \multicolumn{1}{c|}{*}  & 0 & 0 & 0 & 0  \\
		  0 & 0 &  \multicolumn{1}{|c}{0}    & \multicolumn{1}{c|}{\small1} & 0 & 0 & 0 & 0 \\ \cline{3-6}
		  0 & \multicolumn{1}{c}{\cellcolor[gray]{0.8} 0} & 0 & 0 & \multicolumn{1}{|c}{\cellcolor[gray]{0.8} \small1} & \multicolumn{1}{c|}{*} & 0 & 0 \\
		  0 & 0 & 0 & 0 &  \multicolumn{1}{|c}{0}    & \multicolumn{1}{c|}{\small1} & 0 & 0 \\ \cline{5-8}
		  0 & 0 & 0 & 0 & 0 & 0 & \multicolumn{1}{|c}{1} & \multicolumn{1}{c|}{*} \\
		  0 & 0 & 0 & 0 & 0 & 0 & \multicolumn{1}{|c}{0} & \multicolumn{1}{c|}{1} \\ \cline{7-8}
	\end{array}
      \right)\,.
  \end{align*}

  Note that the same matrix $\VVVV1$ has to be applied to all matrices (since the encoder is shared by all users).
  We decompose the resulting matrices (after multiplying them by $\VVVV1$) according to the QR decomposition,
  resulting in unitary matrices $\UUUU{i}{1}$ such that:

  \begin{align} \label{eq:big_btb}
  \nonumber
      &T_i^{(1)} = \UUUU{i}1 \AAA_i \VVVV1  \\
      &\quad = \left(
	\begin{array}{ccccccccc} \cline{1-2}
		\multicolumn{1}{|c}{\smallr_1^i} & \multicolumn{1}{c|}{*} & 0 & 0 & 0 & 0 & 0 & 0 \\
		\multicolumn{1}{|c}{0}    &  \multicolumn{1}{c|}{\cellcolor[gray]{0.8}\smallr_2^i} & 0 & 0 & \multicolumn{1}{c}{\cellcolor[gray]{0.8} 0} & 0 & 0 & 0 \\ \cline{1-4}
		  0 & 0 & \multicolumn{1}{|c}{\smallr_1^i} & \multicolumn{1}{c|}{*}  & 0 & 0 & 0 & 0  \\
		  0 & 0 &  \multicolumn{1}{|c}{0}    & \multicolumn{1}{c|}{\smallr_2^i} & 0 & 0 & 0 & 0 \\ \cline{3-6}
		  0 & \multicolumn{1}{c}{\cellcolor[gray]{0.8}0} & 0 & 0 & \multicolumn{1}{|c}{\cellcolor[gray]{0.8}\smallr_1^i} & \multicolumn{1}{c|}{*} & 0 & 0 \\
		  0 & 0 & 0 & 0 &  \multicolumn{1}{|c}{0}    & \multicolumn{1}{c|}{\smallr_2^i} & 0 & 0 \\ \cline{5-8}
		  0 & 0 & 0 & 0 & 0 & 0 & \multicolumn{1}{|c}{r_1^i} & \multicolumn{1}{c|}{*} \\
		  0 & 0 & 0 & 0 & 0 & 0 & \multicolumn{1}{|c}{0} & \multicolumn{1}{c|}{r_2^i} \\ \cline{7-8}
	\end{array}
      \right) \,,
  \end{align}
where $r_1^i \, r_2^i = 1$ and $i = 2,3$.

  \underline{Stage 2:}
 In the second stage we apply the 1-GMD decomposition to the matrices $T_2^{(1)}\subm{2}{5}$ and $T_2^{(1)}\subm{4}{7}$.
  In both cases the two-by-two matrices are of the same form:

  \begin{align*}
    \Ul22\left(
    \begin{array}{cc}
      r_2^2 & 0 \\
      0	    & r_1^2 \\
    \end{array}
    \right)
    \Vl2
    =
    \left(
    \begin{array}{cc}
      1 & * \\
      0	    & 1 \\
    \end{array}
    \right)
   \,.
  \end{align*}

 Now note that the matrix corresponding to these elements in $T_1^{(1)}$ have the identity matrix form $I_2$.
Thus, $\Vl2$ on the right and $\left( \Vl2 \right)^\dagger$ on the left result in the identity matrix:
  \begin{align*}
    \left(
    \begin{array}{cc}
      1 & 0 \\
      0	& 1 \\
    \end{array}
    \right)
    = \left( \Vl2 \right)^\dagger
    \left(
    \begin{array}{cc}
      1 & 0 \\
      0	& 1 \\
    \end{array}
    \right)
    \Vl2 \,.
  \end{align*}
  For the third matrix, we apply the QR decomposition with $\UUUU32$ (assuming no special structure).
\noindent
Define:

\allowdisplaybreaks{

  \begin{align*}
    \UUUU22 &\triangleq \emb{8}{\Ul22}\left(\subm{2}{5}\subm{4}{7}\right)\\
    \VVVV2 &\triangleq \emb{8}{\Vl2}\left(\subm{2}{5}\subm{4}{7}\right)
    \,.
  \end{align*}

  Thus, we attain the following matrices after the completion of the second stage:

 \begin{align*}
  \nonumber
      &T_2^{(2)} = \UUUU22 T_2^{(1)} \VVVV2  \\
      &\quad = \left(
	\begin{array}{ccccccccc}
		  \smallr_1^2 & * & 0 & 0 & * & 0 & 0 & 0 \\
		  0 &  \multicolumn{1}{c}{\cellcolor[gray]{0.8} 1} & 0 & 0 & \multicolumn{1}{c}{\cellcolor[gray]{0.8} *} & * & 0 & 0 \\
		  0 & 0 & \smallr_1^2 & *  & 0 & 0 & * & 0  \\
		  0 & 0 & 0 & \multicolumn{1}{c}{\cellcolor[gray]{0.5} 1} & 0 & 0 & \multicolumn{1}{c}{\cellcolor[gray]{0.5} *} & * \\
		  0 & \multicolumn{1}{c}{\cellcolor[gray]{0.8} 0} & 0 & 0 & \multicolumn{1}{c}{\cellcolor[gray]{0.8 }1} & * & 0 & 0 \\
		  0 & 0 & 0 & 0 & 0 & \smallr_2^2 & 0 & 0 \\
		  0 & 0 & 0 & \multicolumn{1}{c}{\cellcolor[gray]{0.5} 0} & 0 & 0 & \multicolumn{1}{c}{\cellcolor[gray]{0.5 }1} & * \\
		  0 & 0 & 0 & 0 & 0 & 0 & 0 & r_2^2 \\
	\end{array}
      \right) \,,
\\
  \nonumber
      &T_1^{(2)} = \left( \VVVV2 \right)^\dagger T_1^{(1)} \VVVV2  \\
      &\quad = \left(
	\begin{array}{ccccccccc}
		  1 & * & 0 & 0 & * & 0 & 0 & 0 \\
		  0 &  \multicolumn{1}{c}{\cellcolor[gray]{0.8} 1} & 0 & 0 & \multicolumn{1}{c}{\cellcolor[gray]{0.8} 0} & * & 0 & 0 \\
		  0 & 0 & 1 & *  & 0 & 0 & * & 0  \\
		  0 & 0 & 0 & \multicolumn{1}{c}{\cellcolor[gray]{0.5} 1} & 0 & 0 & \multicolumn{1}{c}{\cellcolor[gray]{0.5} 0} & * \\
		  0 & \multicolumn{1}{c}{\cellcolor[gray]{0.8} 0} & 0 & 0 & \multicolumn{1}{c}{\cellcolor[gray]{0.8} 1} & * & 0 & 0 \\
		  0 & 0 & 0 & 0 & 0 & 1 & 0 & 0 \\
		  0 & 0 & 0 & \multicolumn{1}{c}{\cellcolor[gray]{0.5} 0} & 0 & 0 & \multicolumn{1}{c}{\cellcolor[gray]{0.5} 1} & * \\
		  0 & 0 & 0 & 0 & 0 & 0 & 0 & 1 \\
	\end{array}
      \right) \,,
\\
  \nonumber
      &T_3^{(2)} = \UUUU32 T_3^{(1)} \VVVV2  \\
      &\quad = \left(
	\begin{array}{ccccccccc}
		  \smallr_1^3 & * & 0 & 0 & * & 0 & 0 & 0 \\
		  0 &  \multicolumn{1}{c}{\cellcolor[gray]{0.8} d_2} & 0 & 0 & \multicolumn{1}{c}{\cellcolor[gray]{0.8} *} & * & 0 & 0 \\
		  0 & 0 & \smallr_1^3 & *  & 0 & 0 & * & 0  \\
		  0 & 0 & 0 & \multicolumn{1}{c}{\cellcolor[gray]{0.5} d_2} & 0 & 0 & \multicolumn{1}{c}{\cellcolor[gray]{0.5} *} & * \\
		  0 & \multicolumn{1}{c}{\cellcolor[gray]{0.8} 0} & 0 & 0 & \multicolumn{1}{c}{\cellcolor[gray]{0.8} d_1} & * & 0 & 0 \\
		  0 & 0 & 0 & 0 & 0 & \smallr_2^3 & 0 & 0 \\
		  0 & 0 & 0 & \multicolumn{1}{c}{\cellcolor[gray]{0.5} 0} & 0 & 0 & \multicolumn{1}{c}{\cellcolor[gray]{0.5} d_1} & * \\
		  0 & 0 & 0 & 0 & 0 & 0 & 0 & r_2^3 \\
	\end{array}
      \right) \,,
  \end{align*}
where $d_1 \, d_2 = 1$.

  \underline{Stage 3:}
  Finally, apply the 1-GMD to $\submat{T_3^{(2)}}{4}{5}$:

  \begin{align*}
    \Ul33
    \left(
    \begin{array}{cc}
      d_2 & 0 \\
      0	    & d_1 \\
    \end{array}
    \right)
    \Vl3
    =
    \left(
    \begin{array}{cc}
      1 & * \\
      0	    & 1
    \end{array}
    \right)
    \,.
  \end{align*}

  \noindent Again, note that the corresponding sub-matrices of $T_1^{(3)}$ and $T_2^{(3)}$ are equal to $I_2$.
  Hence, multiplying them by $\Vl3$ on the right and $\left( \Vl3 \right)^\dagger$ on the left, gives rise to the identity matrix $I_2$.
  By defining

\begin{align*}
  \UUUU33 &\triangleq \emb{8}{\Ul33}\left(\subm{4}{5}\right) \\
  \VVVV3 &\triangleq \emb{8}{\Vl3}\left(\subm{4}{5}\right) \\
  \UUUU13 &= \UUUU23\triangleq \left( \VVVV3 \right)^\dagger
\end{align*}

  \noindent we arrive at the following three triangular matrices:

\allowdisplaybreaks{
  \begin{align*}
  \nonumber
      &T_3^{(3)} = \UUUU33 T_3^{(2)} \VVVV3  \\
      &\quad = \left(
	\begin{array}{ccccccccc}
		  \smallr_1^3 & * & 0 & * & * & 0 & 0 & 0 \\
		  0 &  d_2 & 0 & * & * & * & 0 & 0 \\
		  0 & 0 & \smallr_1^3 & *  & * & 0 & * & 0  \\ \cline{4-5}
		  0 & 0 & 0 & \multicolumn{1}{|c}{1} & \multicolumn{1}{c|}{*} & * & * & * \\
		  0 & 0 & 0 & \multicolumn{1}{|c}{0} & \multicolumn{1}{c|}{1} & * & * & * \\ \cline{4-5}
		  0 & 0 & 0 & 0 & 0 & \smallr_2^3 & 0 & 0 \\
		  0 & 0 & 0 & 0  & 0 & 0 & d_1 & * \\
		  0 & 0 & 0 & 0 & 0 & 0 & 0 & r_2^3
	\end{array}
      \right) \,,
\\
  \nonumber
      &T_2^{(3)} = \left( \VVVV3 \right)^\dagger T_2^{(2)} \VVVV3  \\
      &\quad = \left(
	\begin{array}{ccccccccc}
		  \smallr_1^2 & * & 0 & * & * & 0 & 0 & 0 \\
		  0 &  1 & 0 & * & * & * & 0 & 0 \\
		  0 & 0 & \smallr_1^2 & *  & * & 0 & * & 0  \\ \cline{4-5}
		  0 & 0 & 0 & \multicolumn{1}{|c}{1} & \multicolumn{1}{c|}{0} & * & * & * \\
		  0 & 0 & 0 & \multicolumn{1}{|c}{0} & \multicolumn{1}{c|}{1} & * & * & * \\ \cline{4-5}
		  0 & 0 & 0 & 0 & 0 & \smallr_2^2 & 0 & 0 \\
		  0 & 0 & 0 & 0  & 0 & 0 & 1 & * \\
		  0 & 0 & 0 & 0 & 0 & 0 & 0 & r_2^2
	\end{array}
      \right) \,,
\\
  \nonumber
      &T_1^{(3)} = \left( \VVVV3 \right)^\dagger T_1^{(2)} \VVVV3  \\
      &\quad = \left(
	\begin{array}{ccccccccc}
		  1 & * & 0 & * & * & 0 & 0 & 0 \\
		  0 &  1 & 0 & 0 & 0 & * & 0 & 0 \\
		  0 & 0 & 1 & *  & * & 0 & * & 0  \\ \cline{4-5}
		  0 & 0 & 0 & \multicolumn{1}{|c}{1} & \multicolumn{1}{c|}{0} & * & 0 & * \\
		  0 & 0 & 0 & \multicolumn{1}{|c}{0} & \multicolumn{1}{c|}{1} & * & 0 & * \\ \cline{4-5}
		  0 & 0 & 0 & 0 & 0 & 1 & 0 & 0 \\
		  0 & 0 & 0 & 0  & 0 & 0 & 1 & * \\
		  0 & 0 & 0 & 0 & 0 & 0 & 0 & 1
	\end{array}
      \right) \,,
  \end{align*}

  By taking the middle rows and columns (rows and columns 4 and 5) we achieve the desired decomposition with diagonal elements equaling to 1 in all three triangular matrices simultaneously.\footnote{Over the diagonal elements which equal 1, we transmit using SISO codes, in our proposed scheme, whereas we make no use of the remaining elements as they may take arbitrary values.} Formally we do so by defining the next matrix which is composed of rows 4 and 5 of the identity matrix $I_8$,

  \begin {align*}
  \OOO^{\dagger}=\left(
        \begin{array}{cccccccc}
          0 & 0 & 0 & 1 & 0 & 0 & 0 & 0 \\
          0 & 0 & 0 & 0 & 1 & 0 & 0 & 0 \\
        \end{array}
      \right)
    \,,
  \end {align*}
\noindent
  and calculating

  \begin{align*}
  \nonumber
      \OOO^{\dagger} T_{i}^{(3)} \OOO
                = \left(
                  \begin{array}{cc}
                    1 & * \\
                    0 & 1 \\
                  \end{array}
                \right)
       \,.
  \end{align*}

  Thus, the total matrices to be applied are

\begin{align*}
  \mV &= \mV^{(1)} \mV^{(2)} \mV^{(3)} \OOO \\
  \UUU_1 &= \mU_1^{(1)}\VVVV2\VVVV3\OOO\\
  \UUU_2 &= \mU_2^{(1)}\mU_2^{(2)}\VVVV3\OOO\\
  \UUU_3 &= \mU_3^{(1)}\mU_3^{(2)}\mU_3^{(3)}\OOO \,.
\end{align*}

\section{Discussion}
\label{s:Discussion}

According to \thmref{thm:n_n_asymptotical},
even when the number of users $K$ is not very large, a large number of channel uses need to by combined and processed together
in order to approach the capacity.
We demonstrate this phenomenon by considering the example of the introduction \eqref{eq:example_ChannelMatrices} where the gains $\alpha$ and $\beta$ and the power constraint $P$ satisfy \eqref{eq:example_IndividualCapacities}, i.e., that the individual capacities are equal.

For this case, the best achievable rate of the existing practical schemes, described in the introduction, is single-stream beamforming,
which is therefore taken to be the \emph{benchmark} which we aim to improve.
Nonetheless, this scheme does not utilize the two degrees of freedom offered by the first channel $H_1$.
This becomes more significant in the high SNR regime, in which this benchmark rate achieves only half of the available
(multicast) capacity \eqref{eq:MIMO_BC_capacity}.
For the proposed scheme in this paper, we provide in \tableref{table:PortionOfCapacityInfiniteSNR}
the number of channel uses needed to be combined and processed together
to achieve given portions of the capacity, where for comparison,
we present in the table the benchmark rate and the rate achieved by time-sharing between the users,
for the case of $P \rightarrow \infty$ and the case in which the individual capacities~\eqref{eq:example_IndividualCapacities} equal to
$C^\text{p2p} = 10 \left[ \frac{ \text{bits} }{ \text{channel use} } \right]$.\footnote{Both, in the benchmark and the proposed scheme in this paper, we assume that the scalar codes used are capacity-achieving.}

\begin{table}[h]
  \centering
  \begin{tabular}{||c||c|c|c|c|c|c|c|c|c||}
    \hline
    \% Capacity    & 33 & 37 & 50 & 60 & 67 & 75 & 80 & 90 \\ \hline
      Channel uses for GMD &  5 &  5 & 6  & 8  & 9  & 12 & 15 & 30 \\ \hline
      Channel uses for JET &  2 &  2 & 2  & 3  & 3  & 4 & 5 & 10 \\
  \hline
  \end{tabular}
  \captionsetup{justification=justified, font=small}
  \caption{Number of channel uses, when using $K$-GMD and $K$-JET, processed together to achieve a given portion of the capacity.
  For $P \rightarrow \infty$, time-sharing between the users achieves $33\%$ of the capacity, and the benchmark~--- $50\%$;
  for $C^\text{p2p} = 10 \left[ \frac{\text{bits}}{\text{channel use}} \right]$, time-sharing achieves $37\%$ of the capacity whereas the benchmark~--- $67\%$.}
\label{table:PortionOfCapacityInfiniteSNR}
\end{table}
For more users (larger $K$), the ratio between the benchmark rate and the capacity deteriorates rapidly as $K$ grows large. However,
the number of channel uses needed to achieve a certain percentage of the capacity, using the approach developed in this paper, grows rapidly.
Yet, based on numerical evidence, we believe that the number of required channel augmentations
can be reduced.
Furthermore, for special families of MIMO channels, very significant reduction is possible,
as demonstrated in \cite{JET:Permuted_ISIT2012}.


\end{document}